\documentclass[5p,twocolumn]{elsarticle}
\usepackage[USenglish]{babel}
\usepackage[T1]{fontenc}
\usepackage[utf8]{inputenc}
\usepackage{amssymb} 
\usepackage{amsmath}
\usepackage{amsthm}
\usepackage{amsfonts}
\usepackage{siunitx}
\usepackage{booktabs}

\begin{document}


\title{Commissioning results and electron beam characterization at the S-band photoinjector at SINBAD-ARES}

\author[desy]{E.~Panofski}
\ead{eva.panofski@desy.de}

\author[desy]{R.W.~Assmann}
\author[desy]{F.~Burkart}
\author[mol]{U.~Dorda}
\author[desy]{L.~Genovese}
\author[desy]{F.~Jafarinia}
\author[desy]{S.M.~Jaster-Merz}
\author[desy]{M.~Kellermeier}
\author[desy]{W.~Kuropka}
\author[desy]{F.~Lemery}
\author[xfel]{B.~Marchetti}
\author[brook]{D.~Marx}
\author[desy]{F.~Mayet}
\author[desy]{T.~Vinatier}
\author[desy]{S.~Yamin}

\address[desy]{Deutsches Elektronen-Synchrotron DESY,
Notkestraße 85, 22607 Hamburg, Germany}
\address[mol]{SCK CEN,
Boeretang 262, 2400 Mol, Belgium}
\address[xfel]{European XFEL GmbH,
Holzkoppel 4, 22869 Schenefeld, Germany}
\address[brook]{Brookhaven National Laboratory, 2 Center St, Upton, NY 11973, USA}

\begin{abstract}
Over the last years, the generation and acceleration of ultra-short, high quality electron beams has attracted more and more interest in accelerator science. Electron bunches with these properties are necessary to operate and test novel diagnostics and advanced high gradient accelerating schemes such as plasma accelerators or dielectric laser accelerators. Furthermore, several medical and industrial applications require high-brightness electron beams. The dedicated R\&D facility ARES at DESY will provide such probe beams in the upcoming years. After the setup of the normal-conducting RF photoinjector and linear accelerating structures, ARES successfully started the beam commissioning of the RF gun. This paper gives an overview of the ARES photoinjector setup and summarizes the results of the gun commissioning process. The quality of the first generated electron beams is characterized in terms of charge, momentum, momentum spread and beam size. Additionally, the dependencies of the beam parameters on RF settings are investigated. All measurement results of the characterized beams fulfill the requirements to operate the ARES linac with this RF photoinjector.
\end{abstract}

\begin{keyword}
 ARES \sep S-band photoinjector \sep beam commissioning \sep SINBAD \sep DESY
\end{keyword}

\maketitle



\section{Introduction - The ARES Linac at DESY (Overview and Status)}
SINBAD (Short INnovative Bunches and Accelerators at DESY) \cite{DORDA2018} is an accelerator R\&D platform in the former DORIS tunnel at DESY, Hamburg. Its goal is to demonstrate the generation and acceleration of ultrashort electron bunches as well as to test advanced acceleration techniques, such as Laser driven plasma Wake-Field Acceleration (LWFA), Dielectric Laser Acceleration (DLA) and THz-driven acceleration, in multiple independent experiments. 

ARES (Accelerator Research Experiment at SINBAD) \cite{Marchetti2020,Marchetti2018} at SINBAD represents one of these experiments. The linear RF accelerator with a target energy of 100-155\,MeV is currently in the construction and commissioning phase. The facility will provide low-charge, remarkably short electron probe bunches with excellent arrival-time stability \cite{Panofski_2019}. The design parameters of the electron beam are summarized in Table~\ref{tab:ARESparam}. 
\begin{table}[hbt]
   \centering
   \caption{Final electron beam parameters planned at ARES.}
   \vspace{0.2cm}
   \begin{tabular}{ll}
      \toprule
      Repetition Rate [Hz] & 1 - 50\\
			Bunch Charge [pC] & 0.5 - 30\\
			Beam Energy [MeV]  &  54 - 155\\
			Norm. Emittance (rms) [$\mu$m rad]  &  0.1 - 1\\
      Bunch Length (rms) [fs] & 0.2 - 10\\
      Arrival time jitter (rms) [fs]  &  <10\\
      \bottomrule
   \end{tabular}
   \label{tab:ARESparam}
\end{table}
\begin{figure*}[!htb]
   \centering
   \includegraphics*[width=1.0\textwidth]{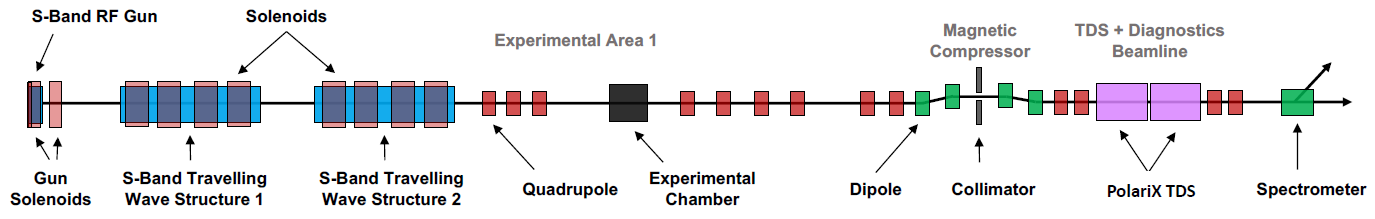}
   \caption{Future layout of the ARES linac at SINBAD.}
   \label{fig:ARES_linac}
\end{figure*}

The SINBAD-ARES setup hosts a normal conducting RF photoinjector generating a low charge electron beam, which is afterwards accelerated by an S-band linac section. The linac, together with a magnetic chicane including a slit collimator, will allow the generation of ultra-short bunches in the fs to sub-fs regime based on the techniques of velocity bunching and magnetic bunch compression \cite{Zhu2016}. The high brightness beam will then serve as a test bench for novel diagnostic devices such as the PolariX TDS, an advanced modular X-band transverse deflecting structure (TDS) system having the new feature of providing variable polarization of the deflecting force \cite{Marx2019,Craievich2020,Marchetti2021}. Furthermore, the ARES accelerator has the potential to provide a test beam for next-generation compact acceleration schemes in the future \cite{Marchetti2016,ZhuPHD,MayetPHD}. ARES will also contribute as a test facility for machine learning for autonomous accelerators, a project of the Helmholtz Artificial Intelligence Cooperation Unit (Helmholtz AI) \cite{AI2020}. 

While SINBAD-ARES is a DESY project, access will be possible to external researchers via the ARIES transnational access program free of charge \cite{Aries2020}. This EU-funded Integrating Activity project aims to develop European particle accelerator infrastructures and includes the task to open research facilities to external users.

It is noteworthy that the ARES linac is sharing several diagnostics components, the control architecture and beam physics challenges of ultra-short, high density electron beams with the AXSIS ERC Synergy Grant project \cite{Matlis2018}. The AXSIS construction is nearing completion in a neighboring space in the same SINBAD facility where ARES is located. Part of the work reported here will be therefore used for training beam instrumentation and procedures that will also be relied on in the upcoming AXSIS beam commissioning, thus minimizing the required time.

Figure~\ref{fig:ARES_linac} illustrates the planned layout of the ARES linac at SINBAD. 
The setup of the SINBAD-ARES facility is proceeding in stages. In a first step, the RF photoinjector including a diagnostic line was completely installed in the ARES tunnel area. The photocathode drive laser system was prepared as well.  After RF conditioning of the gun cavity, a first electron beam was generated and detected at the end of October 2019. This important milestone in the ARES project marked the beginning of the RF photoinjector commissioning and first electron beam characterization.

This publication gives a detailed overview of the accelerator setup in the ARES gun area followed by the results of the RF gun conditioning (Sections 2 and 3). Afterwards, the systematic characterization of the electron beam is presented in Section 4. The following section introduces a beam-based alignment tool for the gun solenoid magnet. Finally, the dark current is described in Section 6.


\section{Overview of the ARES Photoinjector}
\begin{figure}[!htb]
   \centering
   \includegraphics*[width=0.43\textwidth]{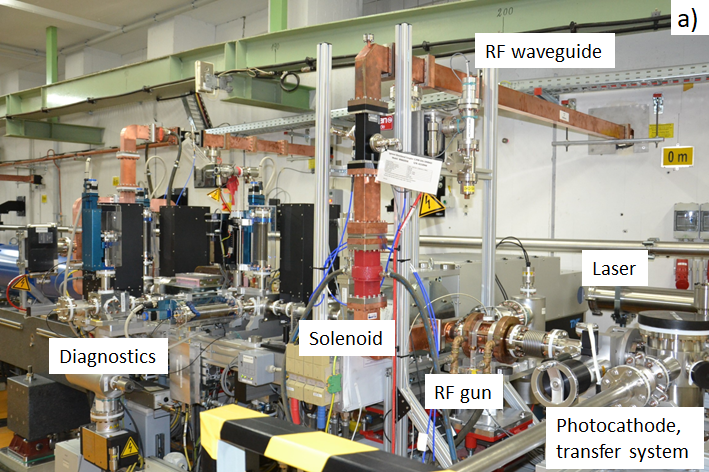}
	 \includegraphics*[width=0.47\textwidth]{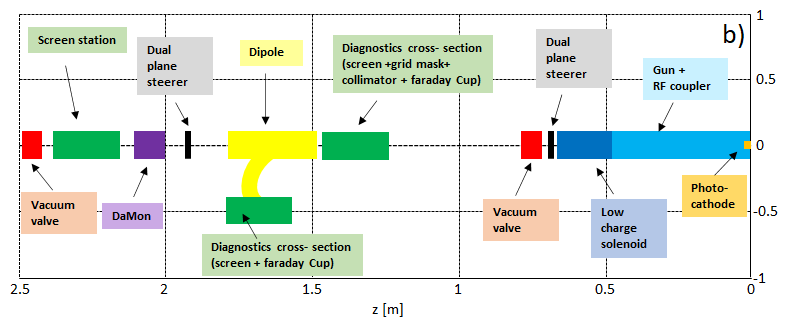}
   \caption{Layout of the ARES photoinjector.
a$)$ Setup of the ARES photoinjector in the SINBAD tunnel including the RF gun with RF waveguides, magnets and a diagnostic line.
b$)$ Schematic overview of the ARES gun section with gun, magnets and diagnostics.
}
   \label{fig:gun_layout}
\end{figure}

Figure~\ref{fig:gun_layout} shows the setup of the ARES gun area in the SINBAD tunnel in January 2020 including infrastructure and hardware \cite{Panofski_2019}. 
The 1.5\,cell, normal-conducting, S-band RF cavity made of copper represents the central part of the setup. The RF is transferred from the klystron to the cavity via a waveguide system that is separated from the cavity itself by a double-coated vacuum window. The electron beam is generated at the photocathode made of a metallic or semiconductive material. The photocathode transfer system allows the insertion of the cathode plug from a cathode box in the back plane of the gun cavity without breaking the vacuum. Electrons are produced via the photoelectric effect \cite{Spicer1993}, therefore, the photocathode is illuminated by the drive laser. At ARES, the laser lab is located outside of the tunnel. The laser is guided through a beampipe under vacuum conditions to the cathode. The chosen cathode materials, molybdenum Mo for the first beam, cesium-telluride Cs$_{2}$Te for the beam commissioning, require UV laser light (257\,nm) for electron generation with maximum efficiency \cite{Dowell2010}. Automated harmonic generators which are integrated in the Pharos laser produce the UV light using the 4$^{th}$ harmonic of the near-infrared photocathode laser \cite{LightConversion2020}. The synchronization between photocathode laser and RF source driving the RF gun of the ARES linac is on the femtosecond level in order to fully utilize the potential of ultra-short electron bunches while probing the novel acceleration techniques \cite{Titberidze2019}. The electron beam is immediately accelerated in the RF gun cavity up to 5\,MeV in order to counteract space charge effects.

Several magnets are available to control the electron beam. A solenoid magnet right after the gun cavity exit focusses the beam in both transverse directions simultaneously. Two dual plane steerer pairs serve for trajectory corrections in the vertical and horizontal directions. A spectrometer dipole guides the electron beam to the dispersive section, allowing momentum spectrum measurements. The accelerator is supplemented with hardware diagnostics. Beam position and beam size measurements can be done at two screen stations in the forward direction and one in the dispersive arm \cite{Wiebers2013}. Beside GAGG (gadolinium aluminium gallium garnet) and LYSO (Lutetium-based scintillator) screens \cite{Kube2019}, several grids are available in the first screen station after the gun cavity for the evaluation of the transverse emittance \cite{Marx2018STAB,MarxPHD}. Two Faraday Cups and one cavity based, and hence non-destructive, charge monitor (DaMon) enable bunch charge and dark current measurements \cite{Lipka2011,Lipka2013}. The Faraday Cup after the spectrometer dipole is also used as a beam dump.

The hardware and diagnostics at ARES are operated by a control system based on DOOCS \cite{Goloborodko1998}. A magnet middle layer server enables a desired magnetic field or transverse kick angle to be specified in addition to adjusting the input current \cite{Froehlich2015}. Furthermore, proper machine settings for different applications can be saved with the Sequencer program. Long-term data acquisition, even for parameters following the repetition rate of up to 50\,Hz, is done with a DAQ server used throughout DESY \cite{Agababyan2005}. Leveraging the numerous possibilities to interface with the control system (MATLAB, Python, Java,C/C++) the operators contributed software both for commissioning purposes and day-to-day operation of the machine.


\section{RF Conditioning of the Gun Cavity and RF Stability}
\begin{figure}[!htb]
   \centering
   \includegraphics*[width=0.4\textwidth]{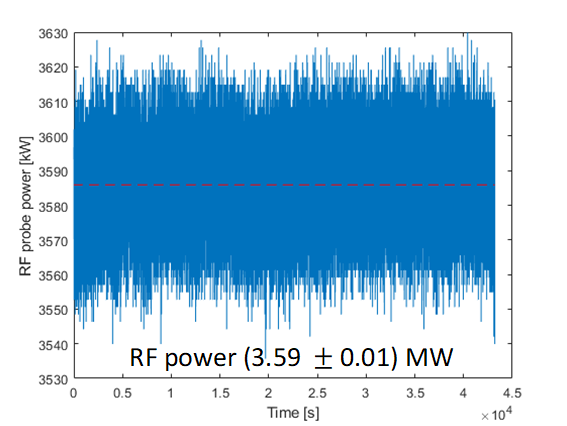}
   \caption{Stability measurement of the RF probe power in the ARES gun over 12 hours.}
   \label{fig:RF_stability}
\end{figure}

As a first step during the accelerator commissioning, the gun cavity was fully conditioned. Currently, a maximum RF power of 3.7\,MW with an RF pulse length of 4.0\,$\mu$s has been achieved during operation. Measurements over 12\,hours confirm a stable RF power in the gun cavity with fluctuations less than 10\,kW 
[see Fig.~\ref{fig:RF_stability}]. 

\begin{figure}[!htb]
   \centering
   \includegraphics*[width=0.47\textwidth]{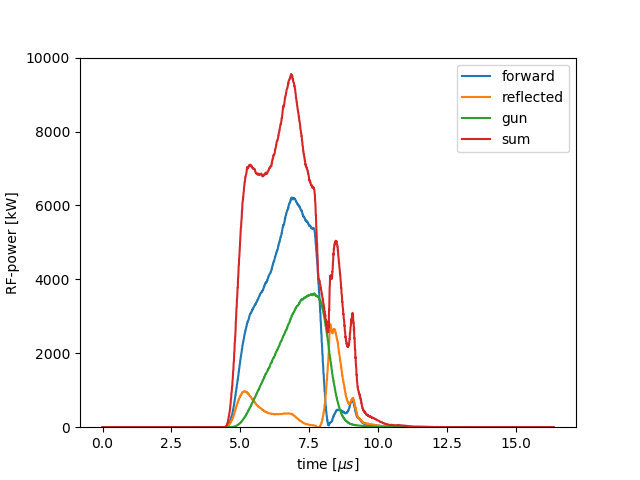}
   \caption{RF power signals in the gun cavity: Displayed are the forward (blue), probe (green) and reflected (yellow) power readout. The red curve represents the sum power given by the quadratic sum of forward and reflected power.}
   \label{fig:RFpower}
\end{figure}

The final goal is to run the ARES gun at an RF probe power of 6 MW that will allow acceleration of the electron beam to more than 5\,MeV. A higher beam energy at the RF gun exit always helps to better control space charge effects since the transversely defocusing space charge force in a drift scales with 1/$\gamma$, with $\gamma$ being the Lorentz factor \cite{Reiser}. The current limit to the maximum RF power coupled to the gun cavity is mainly defined by the maximum power limit of the klystron, which is given by the quadratic sum of the forward and reflected power that can theoretically reach the modulator. Due to a high amount of reflection, the sum power comes close to the hard limit of the modulator of 11\,MW. Consequently, the forward and therefore the probe power in the gun cavity cannot be further increased. Figure~\ref{fig:RFpower} illustrates this fact by showing the forward (blue curve), reflected (yellow curve) and the resulting probe power signal (green curve) measured in the RF gun. The red curve gives the sum power that nearly reaches the 11 MW limit of the klystron window. In the future, a circulator will be installed, which will filter out the reflected part of the power signal, providing a way toward increased probe power in the gun. 

\section{Electron Beam Characterization}
After the first electrons were detected at the end of October~2019, the electron beam was systematically characterized. 

\subsection{Bunch charge}
We started with bunch charge measurements and the dependency of the generated charge on laser pulse energy and RF parameters. At ARES, the charge can be detected using a Faraday Cup and a DaMon in the forward direction of the beamline, as well as an additional Faraday Cup in the dispersive section. A bunch charge of (16.20$\pm$0.03)\,pC at the Faraday Cups and (16.70$\pm$0.04)\,pC at the DaMon was measured at the current maximum RF power (3.7\,MW) and an RF phase -20\,deg. The RF phase is always defined relative to zero-crossing (total RF phase +2\,deg) where no net acceleration occurs and hence, no bunch charge is detected. The charge values that are achieved with the two different measurement techniques agree well.

Following the theory of the photoelectric effect, the parameters which mainly impact the number of electrons extracted from the cathode, i.e. the bunch charge $q_{b}$ are: the pulse energy of the drive laser $E_{laser}$ and thus the number of incoming photons, and the quality of the photocathode i.e. the quantum efficiency $QE$. The latter is mainly defined by the cathode material itself (metal or semiconductor), the surface roughness and the quality of coating of the cathode plug. The number describes the ratio of emitted electrons to incoming laser photons. 
\begin{equation}
  q_{b} = \frac{eE_{laser}}{E_{ph}}QE.
  \label{eq:bunchcharge}
\end{equation}
In Formula~\ref{eq:bunchcharge}, $E_{ph}$ describes the photon energy determined by the laser wavelength. In addition, the bunch charge can be increased with a higher RF field at the cathode. Due to this so-called Schottky effect the RF field lowers the effective surface potential of the cathode and more electrons can escape the photocathode material \cite{Dowell2009}.
 
\begin{figure}[!htb]
   \centering
   \includegraphics*[width=0.47\textwidth]{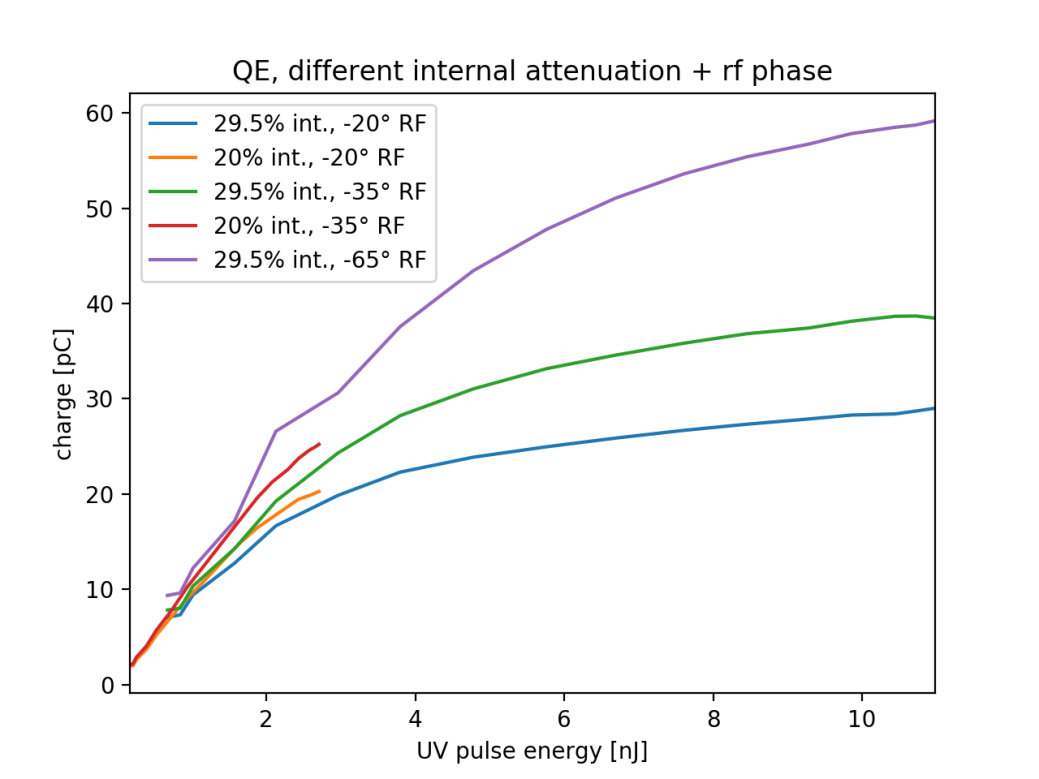}
   \caption{Charge measurement during a scan of the photoinjecor laser pulse energy at the photocathode surface for different laser attenuator positions and RF phases in the gun cavity.}
   \label{fig:QE}
\end{figure}

While the quantum efficiency of the photocathode cannot be varied in a controlled way during a measurement without a cathode exchange, the laser pulse energy and the RF field at the cathode can be modified. Figure~\ref{fig:QE} shows the bunch charge that was measured with the Faraday Cup while the laser pulse energy was continuously changed by using an attenuator on the laser table. An additional internal attenuator reduced the outcoming laser energy to 20\% and 29.5\% respectively in order to limit the emitted bunch charge to tens of pC (see legend in Fig.~\ref{fig:QE}). The measurement was repeated for three different RF phases, -20\,deg relative to zero-crossing, -35\,deg RF phase that corresponds to a maximum beam momentum and -65\,deg RF phase that gives maximum charge output. Increasing the laser energy the bunch charge grows linearly up to a laser energy of about 2
\,nJ. The electron emission is limited by the amount of incoming laser photons and can be described with Formula~\ref{eq:bunchcharge}. For laser energies above 2\,nJ, the charge emission is increasingly space-charge dominated and runs into the space-charge limit (plateau). The electrons experience their own image charge at the cathode surface, which produces a field opposing the applied RF field. The electron emission saturates when the electric field that is triggered by space charge equals the RF field at the surface of the cathode \cite{Bazarov2009}. 
\\
At ARES, a Cs$_{2}$Te photocathode was used for the beam characterization measurements. Based on the charge measurement data illustrated in Figure~\ref{fig:QE} and the theory of photon limited emission (see Formula~\ref{eq:bunchcharge}) the quantum efficiency is calculated as (4.5$\pm$0.2)\% for an RF phase of -20\,deg and (5.1$\pm$0.2)\% for an RF phase of -35\,deg, respectively. The difference between these two numbers is due to the Schottky effect that increases the amount of extracted electrons from the photocathode at lower RF phase (here -35\,deg.) The calculated quantum efficiencies are realistic compared to measurements from other electron sources \cite{Wisniewski2015}. The combination of quantum efficiency and available laser pulse energy is sufficient to operate the linac in the charge range of a few pC to several hundreds of pC extrapolating the data from the measurements shown in Figure~\ref{fig:QE}. During the commissioning phase the charge was limited to tens of pC in order to control the dark current and to protect diagnostics.

\begin{figure}[!htb]
   \centering
   \includegraphics*[width=0.47\textwidth]{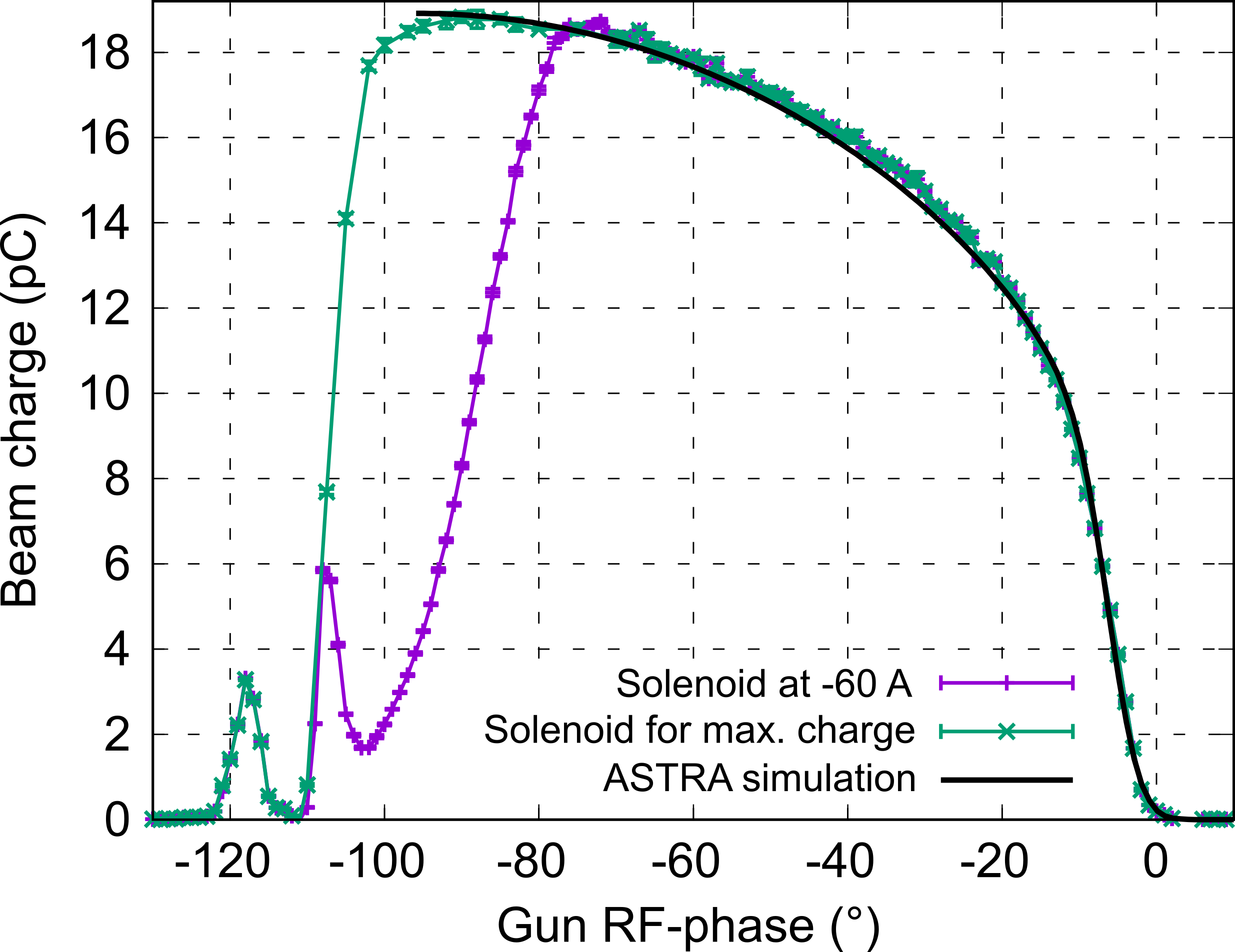}
   \caption{Bunch charge measurement during an RF phase scan (Schottky scan). Purple colored measurement data were achieved at a constant solenoid current of -60\,A. The solenoid current was optimized for maximum charge at the DaMon for the green curve. A fit with an ASTRA simulation allows the duration of bunch emission to be estimated \cite{ASTRA}.}
   \label{fig:Schottky}
\end{figure}

The dependency of the emitted charge on the RF field at the cathode can be illustrated with a Schottky scan. In the experiment, the RF phase has been continuously varied during the charge measurement. Figure~\ref{fig:Schottky} shows the result of two Schottky scans done with the DaMon in the gun area of ARES. A constant RF power of 3.6\,MW was used for a rapid acceleration of the electrons starting from the cathode.  
\\
The purple colored curve in Figure~\ref{fig:Schottky} displays the Schottky scan at a constant solenoid current of -60\,A. The selected solenoid field allows the full amount of charge to be focused to the DaMon in an RF phase range from 0\,deg to around -70\,deg. For smaller RF phases it becomes necessary to adjust the solenoid focusing. This was done in a second measurement run that corresponds to the green data in Figure~\ref{fig:Schottky}.

In general, the Schottky curve that displays the measured bunch charge vs. the RF phase and therefore the RF field at the surface of the photocathode can be explained as follows: Starting from an RF-phase above 0\,deg, the entire cathode laser pulse impacts the cathode when the gun field is decelerating. Hence, no electrons are accelerated and no charge is detected at the DaMon. Below 0\,deg down to -10\,deg an increasingly large part of the laser pulse hits the cathode when the gun field is accelerating. More and more photo-emitted electrons are extracted and the Schottky curve rises. The bunch reaches its target bunch length defined by the laser pulse duration at around -10\,deg. Instead of a constant beam current at RF phases below -10\,deg (plateau), an additional increase can be detected in Figure~\ref{fig:Schottky}. Further electrons escape from the photocathode due to the external RF field lowering the potential barrier. A lowered surface barrier enables non-excited electrons to tunnel into the vacuum and favors the extraction of excited electrons. At around -70\,deg, the beam momentum becomes too low. Consequently, the beam is over-focussed at the DaMon by a constant solenoid current of -60\,A and the measured bunch charge drops in Figure~\ref{fig:Schottky} (purple curve). In the case of a variable solenoid current (green curve) that is adjusted to achieve a maximum charge at the DaMon, a further increase of the bunch charge due to Schottky effect up to 14\,pC at -90\,deg can be observed. Afterwards, the momentum spectrum becomes too wide to be fully transported to the detector, which leads to a decrease in the measured charge also in the case of variable solenoid currents (green curve). The two peaks visible on the left side of the scan (purple curve) are due to electrons travelling back and forth in the gun before exiting and to secondary electrons emitted through cathode back-bombardment. These electrons can have high enough energies (namely close enough to the maximum momentum) to be focused by the gun solenoid and collected by the Faraday cup. Another potential contribution is from very low energy electrons undergoing multiple foci in the gun solenoid field before exiting it.

The linear slope of the Schottky curve at RF phases close to zero can be used to estimate the duration of the electron bunch emission from the photocathode. The curve in Fig.~\ref{fig:Schottky} is calibrated with an ASTRA simulation using a Gaussian laser pulse as input to simulate the photoelectric effect and assuming an instantaneous response from the photocathode \cite{ASTRA}. As a result, an emission duration of (2.5$\pm$0.5)\,ps (RMS) is estimated.

\subsection{Beam momentum and momentum spread}
The momentum of the electron beam after acceleration in the ARES gun has been evaluated using two different methods. The spectrometer dipole in the gun section allows the beam momentum to be determined based on the dipole current that deflects the beam by 90\,deg. The spectrometer dipole was previously calibrated for this measurement.

As an alternative the first steerer magnet together with a screen can be used for a momentum measurement. In this case the momentum can be calculated using the following formula 
\begin{equation}
  pc = \frac{elLc\Delta B}{\Delta x} = \frac{elc\Delta B}{\Delta\theta},
  \label{eq:momentum}
\end{equation}
where $e$ represents the elementary charge and $c$ is the speed of light. $l$ is the steerer magnetic length, $L$ gives the distance between the steerer and imaging screen, $\Delta x$ is the displacement of the beam on the screen, $\Delta B$ describes the variation of the steerer magnetic field and $\Delta\theta$ represents the corresponding kick induced by the steerer. Knowing the calibration of the steerer, i.e. the magnetic length and the relation between steerer current and steerer magnetic field, the beam momentum $pc$ can be retrieved from the measurement of $\Delta x$. This technique of determining the momentum is valid as long as L$\gg$l and no element is placed between steerer and screen which could induce a steerer-dependent kick. These assumptions are valid in the ARES gun setup. Hence the steerer-based momentum measurement is routinely used as a fast alternative to the spectrometer measurement.

Since the orbit correction at ARES is done with dual-plane steerers, the momentum can be determined using a horizontal as well as a vertical deflection. The average of the results from both planes is calculated and afterwards compared with the momentum values obtained from spectrometer measurements. The momentum measurement has been done for 3 different RF power values and a constant RF phase of -20\,deg relative to zero-crossing using both techniques. Table~\ref{tab:Momentum} summarizes the obtained results. 
\begin{table}[hbt]
   \centering
   \caption{Momentum measurement results for different RF powers.}
   \vspace{0.2cm}
   \begin{tabular}{lll}
      \toprule
      Power [MW] & pc steerer [MeV]  &  pc dipole [MeV]\\
      3.70$\pm$0.19 & 4.63$\pm$0.06 & 4.64$\pm$0.04\\
      3.26$\pm$0.16 & 4.39$\pm$0.06 & 4.39$\pm$0.06\\
      2.77$\pm$0.14 & 4.05$\pm$0.05 & 4.09$\pm$0.05\\
      \bottomrule
   \end{tabular}
   \label{tab:Momentum}
\end{table}\\
The measurement data from both methods are very well comparable. Differences are within the error bars.\\
Table~\ref{tab:Momentum} also displays the expected dependency of the beam momentum on the RF power. A higher RF power increases the strength of the accelerating field that allows higher electron beam momenta after the gun cavity.

\begin{figure}[!htb]
   \centering
   \includegraphics*[width=0.47\textwidth]{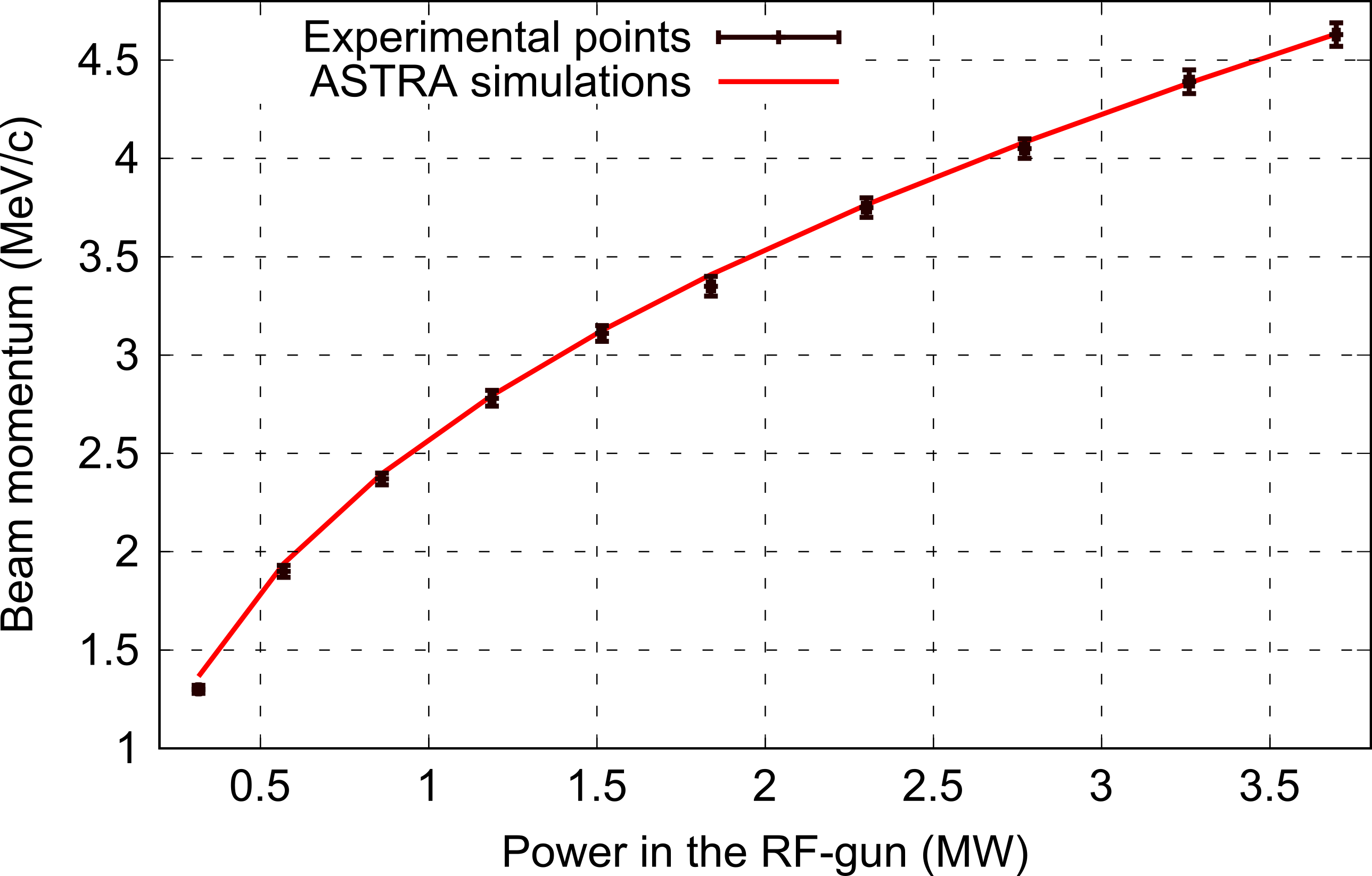}
   \caption{Measured beam momentum vs. RF power. The momentum measurement was done using the first vertical corrector magnet.  The RF phase was fixed at -20\,deg relative to zero-crossing. The measurement points are compared with ASTRA simulations.}
   \label{fig:Momentum_power}
\end{figure}

A measurement of the beam momentum during an RF power scan was done. Figure~\ref{fig:Momentum_power} shows the measurement data and the corresponding ASTRA simulation at a constant RF phase of -20\,deg from the zero-crossing. The accelerating field used in the ATSRA simulations is based on the shunt impedance of the gun and has been determined via the LLRF law  
\begin{equation}
 E(MV/m) = 47.67\sqrt{P(MW)}.
  \label{eq:LLRF}
\end{equation} 
Formula~\ref{eq:LLRF} calculates the accelerating gradient $E$ from the RF probe power $P$ in the gun. The measurement results fit well with the simulation data from ASTRA and verify the LLRF law simultaneously.

\begin{figure}[!htb]
   \centering
   \includegraphics*[width=0.47\textwidth]{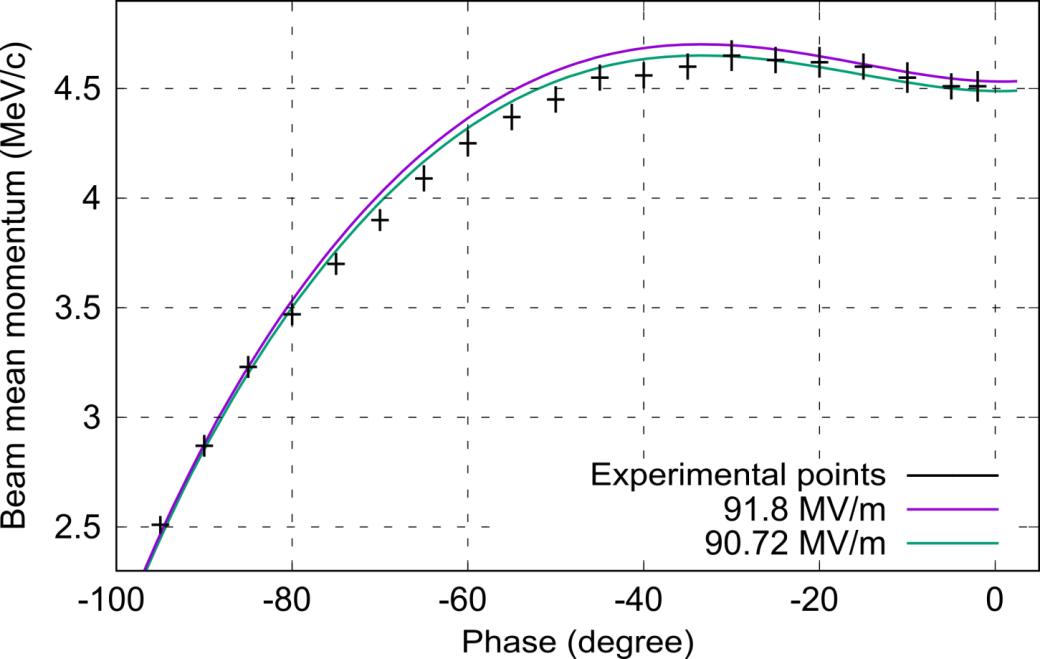}
   \caption{Measured beam momentum vs. RF phase. The acceleration gradient is estimated with around 91\,MV/m from fitting ASTRA simulations.}
   \label{fig:Momentum_phase}
\end{figure}

Besides the RF power, the injection phase of the electron beam in the RF wave impacts the final beam momentum.  An additional scan of the RF phase between -100\,deg and 0\,deg illustrates this fact (see Fig.~\ref{fig:Momentum_phase}). Maximum acceleration at this RF power in the gun (3.7\,MW), and therefore maximum beam momentum, is obtained at around -30\,deg, the so called “on-crest” phase. An acceleration gradient between 90.7\,MV/m and 91.8\,MV/m can be estimated from the ASTRA simulations that fits the measurement data best. An acceleration gradient of 91.7\,MV/m can be estimated from the LLRF law.

Furthermore, the spectrometer dipole was used to determine the momentum spread of the electron beam. The bunch was deflected by 90\,deg (only particles with target momentum) in the dipole and afterwards projected on a screen. Particles with deviating momenta are deflected by angles different from 90\,deg. As a consequence, the horizontal beam width that is recorded on the screen, gives a measure of the momentum spread. At a maximum RF power of 3.7\,MW and an RF phase of -20\,deg from the zero-crossing, an RMS momentum spread of (8.84$\pm$0.09)\,keV/c is determined. This value corresponds to 0.19\% of the beam average momentum at these RF settings. The momentum spread which is obtained in an ASTRA simulation is slightly different at 11.9\,keV/c. However, the ASTRA simulation is based on the emission duration of (2.5$\pm$0.5)\,ps estimated from the Schottky scan as input parameter. The discrepancy in the measured and simulated momentum spread numbers can well be explained with the large error on the emission duration of the beam.

\subsection{Transverse spot size}
\begin{figure}[!htb]
   \centering
   \includegraphics*[width=0.47\textwidth]{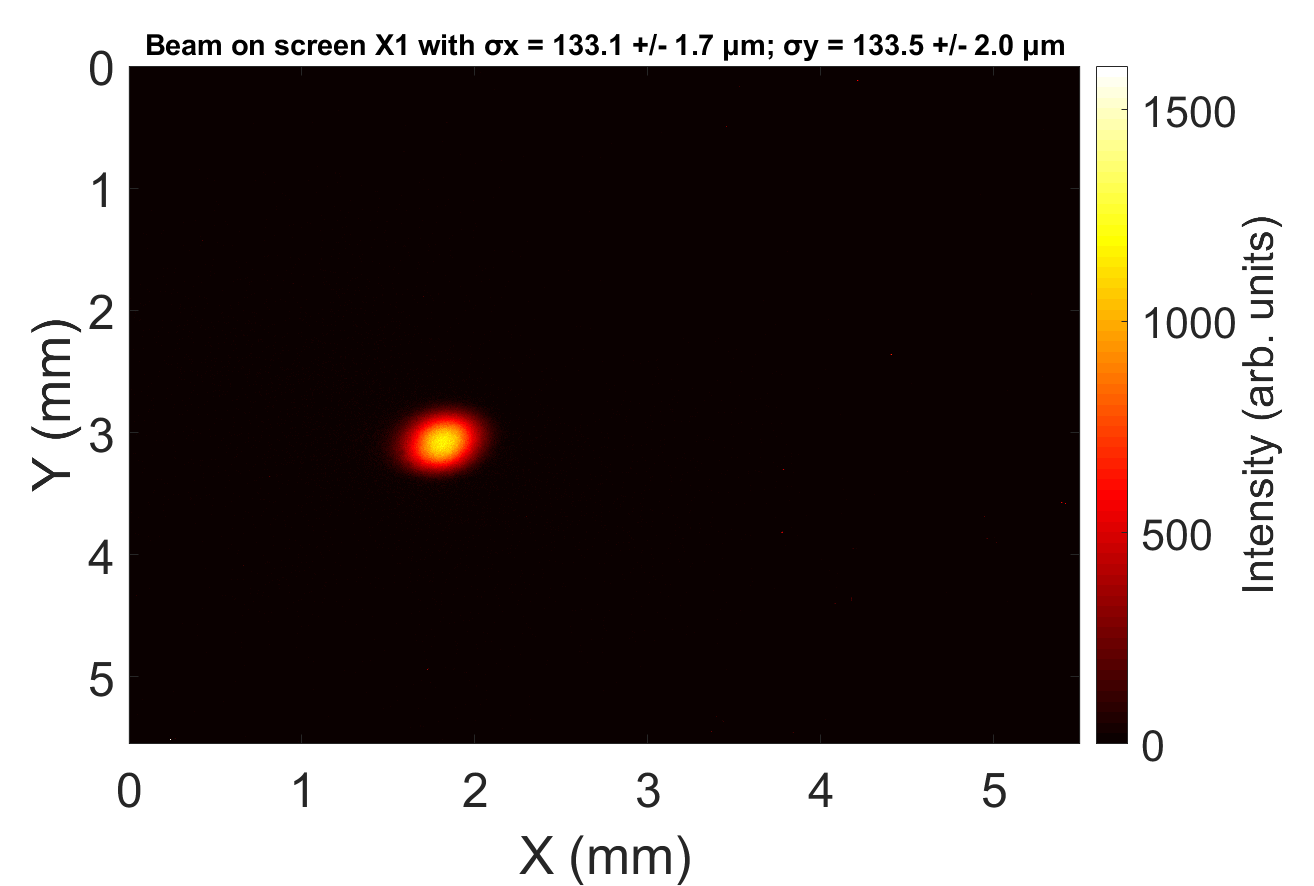}
   \caption{Beam size measurement on the first screen in the gun section.}
   \label{fig:beam_size}
\end{figure}

\begin{figure}[!htb]
   \centering
   \includegraphics*[width=0.23\textwidth]{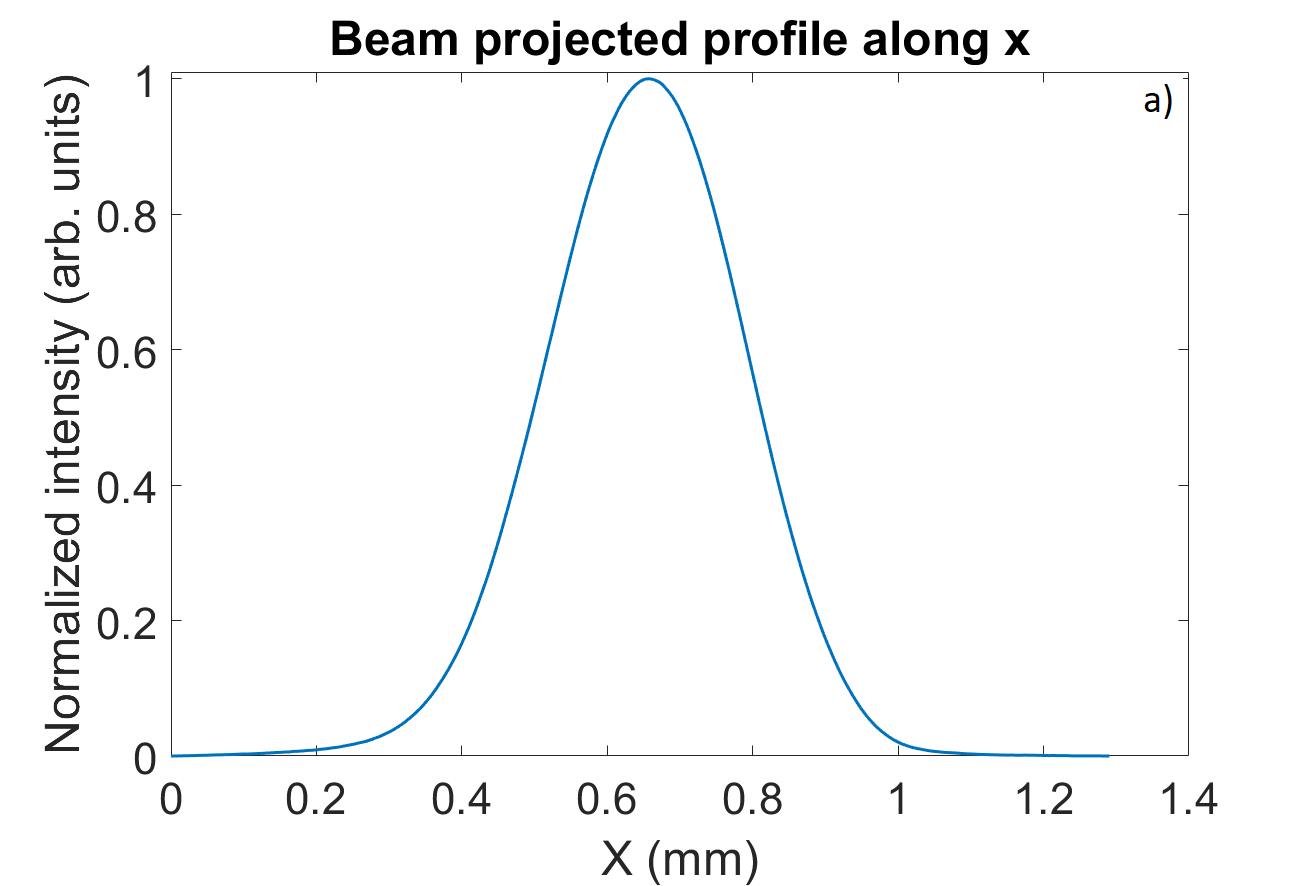}
	 \includegraphics*[width=0.23\textwidth]{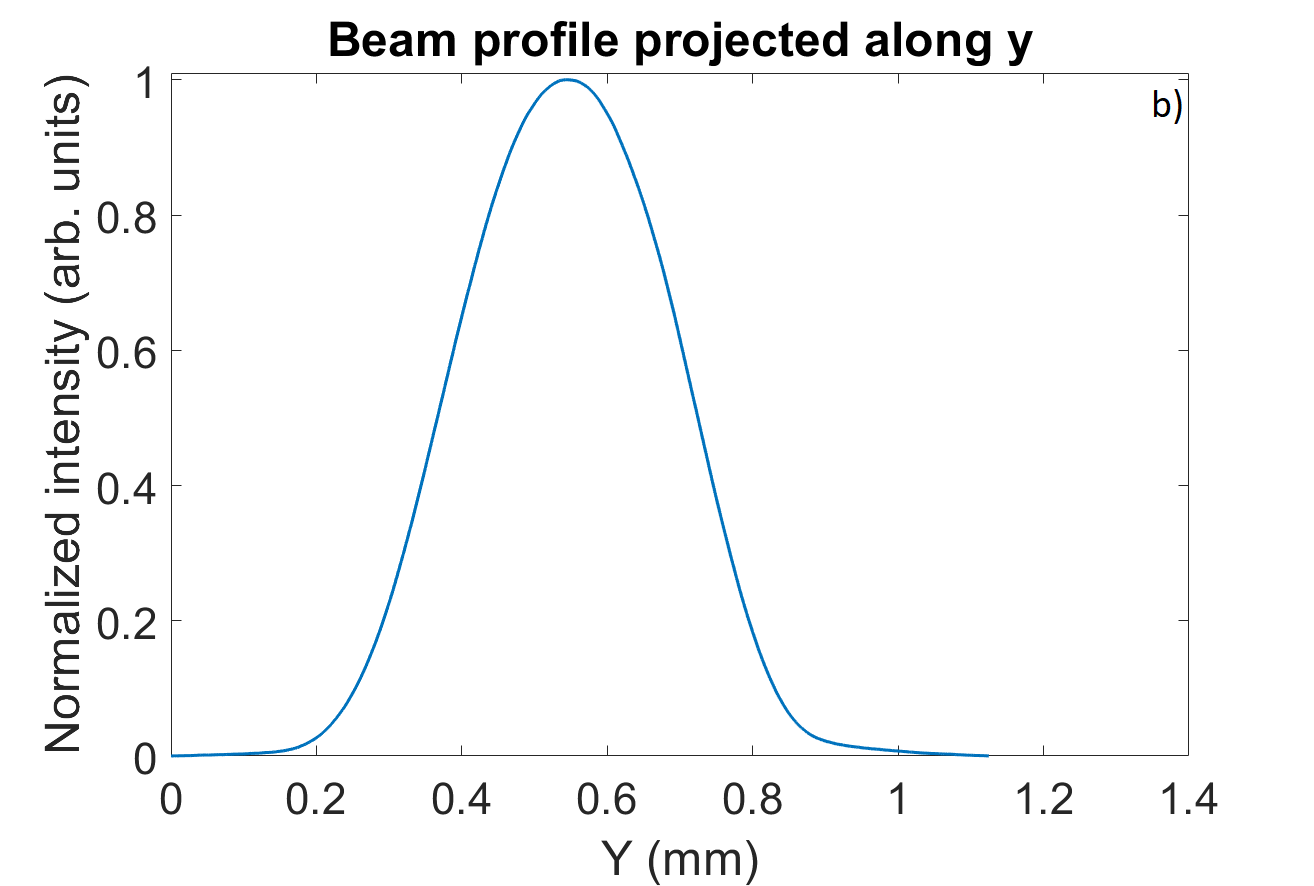}
   \caption{Analysis of the beam profile along x- (a) and y-direction (b).}
   \label{fig:xy_profile}
\end{figure}

The electron beam was projected on the first screen in the ARES beamline, around 1.3\,m behind the photocathode. Figure~\ref{fig:beam_size} displays the corresponding camera image. The RF gun accelerated the (7.5$\pm$0.1)\,pC beam to a momentum of (4.76$\pm$0.07)\,MeV/c before the solenoid focused it on the screen. An RMS beam size of around 133\,$\mu$m was measured. The focus of this measurement was on a perfectly symmetric beam. RMS beam widths of $\sigma_{x}$=(133.1$\pm$1.7)\,$\mu$m and $\sigma_{y}$=(133.5$\pm$2.0)\,$\mu$m in x- and y-direction were obtained with a round laser pulse and focused by the pre-aligned solenoid magnet. The symmetry can be also seen in the beam profiles along x and y that are plotted in Figures~\ref{fig:xy_profile}a) and \ref{fig:xy_profile}b). 
\\
Several different measurement techniques may be used to characterize the transverse emittance including metal grids, a pepper-pot mask and focusing magnets. At the ARES photoinjector the emittance can be evaluated by scanning the magnetic strength of the gun solenoid and measuring the beam size on one of the screens downstream. Since the solenoid magnet was not completely aligned at this point of commissioning not enough measurement data of good quality could be taken. As an alternative a pepper pot and several metal grids are available in the ARES gun section, which allow not only an emittance measurement but the full reconstruction of the 4D phase space. A pepper pot cuts small beamlets out of the electron beam that are afterwards imaged on a screen downstream. The moments can be calculated from the envelope of the beamlet peaks, the beamlet widths and their mean positions on the screen \cite{MarxPHD}. Similarly, when using metal grids, which are better suited for low-charge bunches, the shadow image of the bars on the screen may be analyzed to calculate the moments. 
However, due to the initial solenoid misalignment the pointing of the electron beam must be compensated by operating the spectrometer dipole with a small current. As a consequence, the beam was sheared and the image of the beamlets on the last screen could not be analyzed. Further emittance measurements are planned after the solenoid alignment is fully finished. Measurement tools and analysis software are available. Meanwhile, ASTRA simulations were performed to estimate the transverse emittance for different laser settings. An emittance below 0.5\,mm\,mrad is expected for a 10\,pC electron beam.


\section{Beam-based alignment of the gun solenoid}
The beam-based alignment of the gun solenoid represents an essential step during the RF gun commissioning \cite{Krasilnikov2005}. In order to preserve the low emittance of an electron beam a good alignment of the axis of the focusing element with the beam trajectory is mandatory. 

During our alignment process, the solenoid current was scanned and the corresponding beam centroid position was recorded on the first screen downstream in the beamline. The obtained data were fit using an algorithm based on linear transfer matrices to compute the solenoid misalignment both in terms of offset s and angles \cite{Yamin_2019}. The script is implemented in a Matlab tool. A Matlab GUI visualizes the movement of the beam spot on the screen with changing solenoid current. In addition, the misalignment in the solenoid position, calculated by the Matlab script, is displayed.

\begin{figure}[!htb]
   \centering
   \includegraphics*[width=0.47\textwidth]{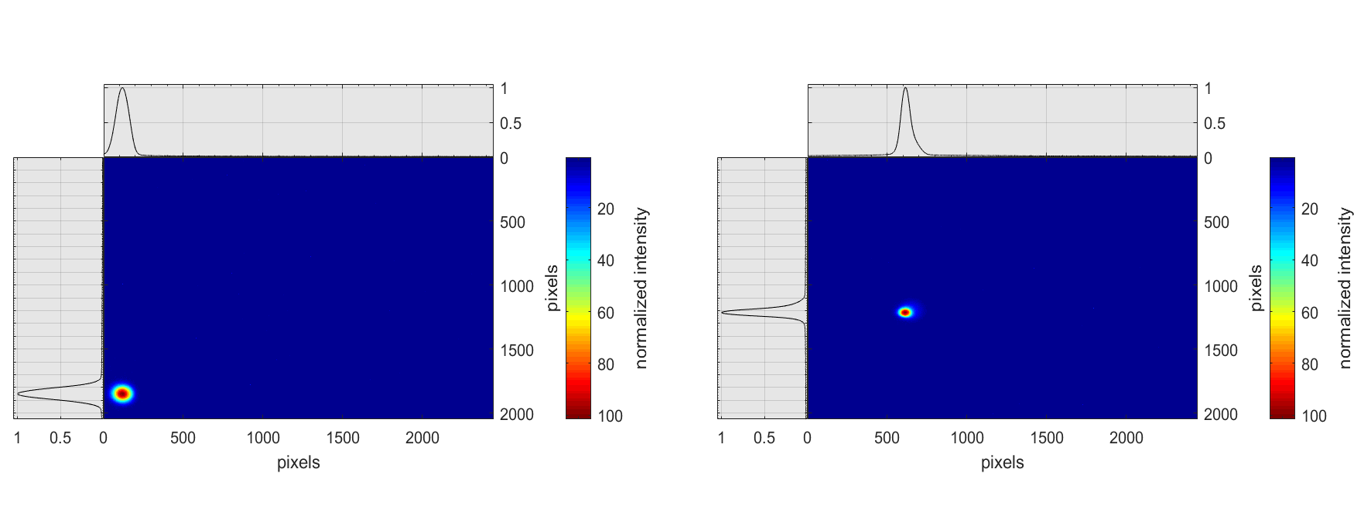}
   \caption{Comparison of the beam position on the first screen before (left) and after (right) the correction
of the solenoid position in the horizontal plane. All other conditions (laser, RF settings) are identical.}
   \label{fig:BBA}
\end{figure}

In the first iteration of the solenoid alignment process, data were taken and analyzed for two different power levels (3.6\,MW and 2.6\,MW) with ++ and +- polarity settings of the double coil gun solenoid respectively. Table~\ref{tab:SolAlig} gives an overview of the calculated solenoid offset in x- |dX| and y-direction |dY| as well as its rotation around the x- |dX$_{rot}$| and y-axis |dY$_{rot}$| averaged over multiple data sets. 
\begin{table}[hbt]
   \centering
   \caption{Results for the solenoid misalignment obtained for both polarity settings (++, +-) and two RF powers (3.6\,MW and 2.6\,MW) from the beam-based alignment tool.}
   \vspace{0.2cm}
   \begin{tabular}{ll}
      \toprule
      Misalignment& Estimated value\\
			|dX| (mm) & 0.81$\pm$0.16\\
			|dX$_{rot}$| (mrad) &  1.33$\pm$0.27\\
			|dY| (mm) &  0.38$\pm$0.08\\
      |dY$_{rot}$| (mrad) & 1.98$\pm$0.40\\
      \bottomrule
   \end{tabular}
   \label{tab:SolAlig}
\end{table}\\
In order to check the validity of the results from the alignment script, a correction of the x-offset was experimentally applied. The solenoid was moved by 0.8\,mm in the horizontal direction. Figure~\ref{fig:BBA} shows a clear improvement in the beam position on the first screen after the beam-based alignment correction. However, it must be noticed that the beam position changed not only in the horizontal but also slightly in the vertical position. It was observed that the movers of the solenoid are not fully decoupled in their horizontal and vertical motion. Furthermore, since the beam position still changes during a solenoid current scan, further iterations are required in the alignment process. It is foreseen to repeat the procedure in order to perfectly align the ARES gun setup.

\section{Dark Current Characterization}
Beside the photo-emitted electrons that form the electron bunch, a significant amount of dark current electrons was observed. The intensity of the most prominent dark current spot exceeds the one of the photo-electron beam. Therefore, a subtraction of the background (including dark current) is required for the analysis of screen images and charge measurement data. 

\begin{figure}[!htb]
   \centering
   \includegraphics*[width=0.47\textwidth]{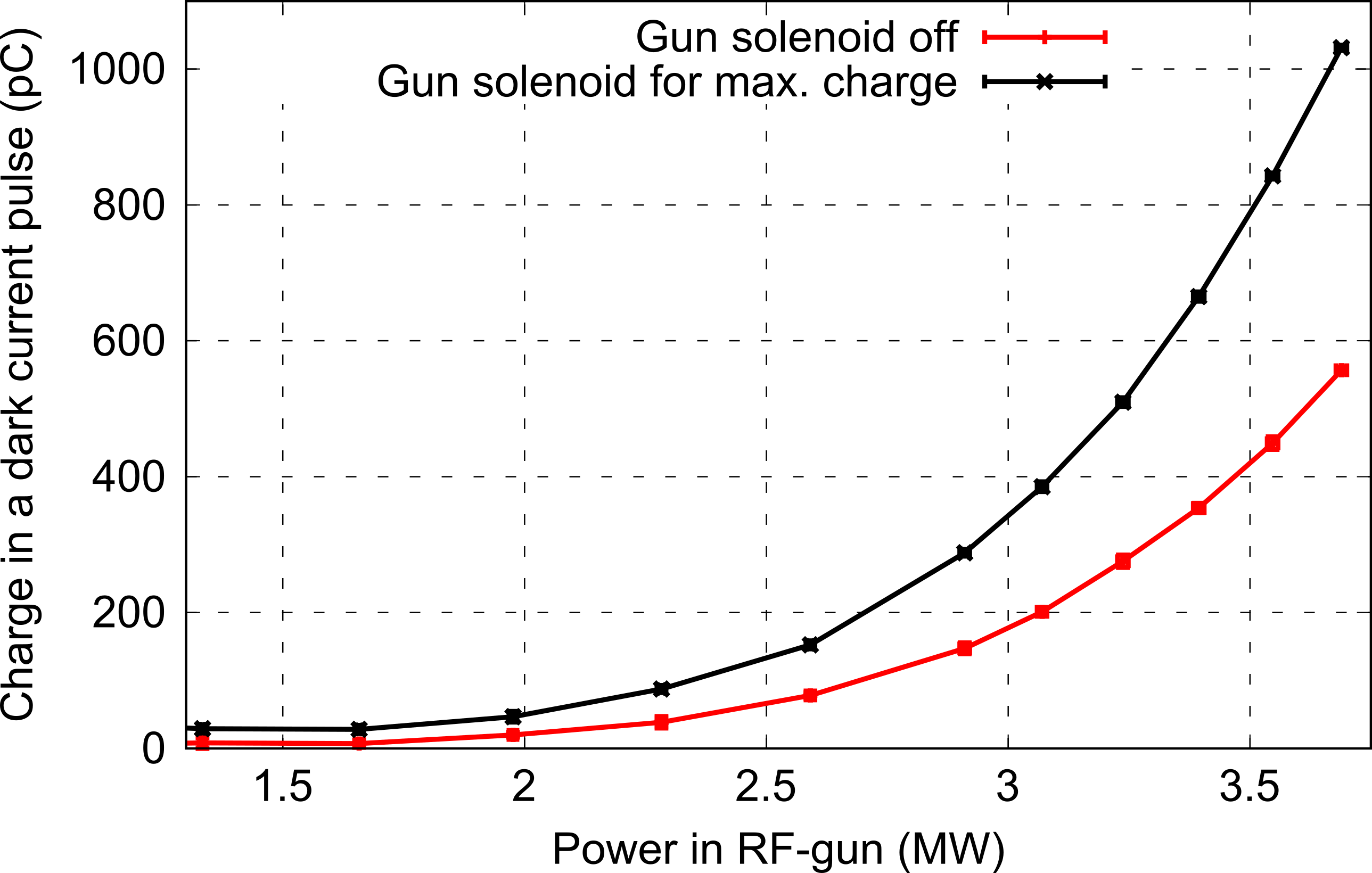}
   \caption{Dark charge measurement during an RF power scan. The data were taken with and without solenoid focusing to the Faraday Cup.}
   \label{fig:DC_power}
\end{figure}

The dependency of the dark current on the RF power is illustrated in Figure~\ref{fig:DC_power}. A Faraday Cup measured the dark current charge (integrated dark current over the RF pulse) while the RF power in the gun cavity was increased. In order to focus the dark current electrons to the Faraday Cup and to cover the charge of most of the dark current spots, a second measurement sequence was performed with the ARES gun solenoid switched on (see Fig.~\ref{fig:DC_power} black colored data). The solenoid current was varied to maximize the measured dark current charge at each RF power. Figure~\ref{fig:DC_power} confirms that the currently installed RF gun cavity with a Cs$_{2}$Te photocathode delivers a dark charge level of around 1\,nC at a maximum RF power of 3.7\,MW and an RF pulse length of 4\,$\mu$s. 

The dark current was further characterized by a momentum measurement using one steerer magnet. The momentum of the most significant dark current spot is (4.56$\pm$0.06)\,MeV/c and, thus, close to the maximum momentum of the electron beam. Therefore, it is assumed that the source(s) of the dark current electrons can be mostly found at the back plane of the gun cavity. Probably, field emitters generate dark current, which also explains the increase in the charge while the RF power, and therefore the field at the cavity surface, is maximized.

Further investigations of the dark current are planned and strategies to lower the dark current will be tested in order to avoid an activation of the machine and to enable high quality, low charge (sub-pC range) user experiments at ARES. The gun was exchanged with an upgraded version of the ARES gun cavity with a different geometry in spring 2020. This new gun is especially designed for SINBAD ARES. An additional solenoid can be installed around the cavity that will allow advanced emittance compensation in a high charge mode. A characterization of the electron beam as well as of the dark current generated from the new gun is ongoing. 


\section{Conclusion and Outlook}
This paper summarized the RF photoinjector commissioning results for ARES gun 1. The RF conditioning of the gun cavity was finished and the solenoid magnet was initially aligned. Afterwards, we characterized the beam quality in terms of charge, momentum, momentum spread and beam size. Furthermore, we investigated the dependencies of these parameters on RF settings (power, phase). All measured beam parameters fulfill the requirements to operate the ARES linac with this electron source. However, the high amount of dark current generated in the RF photoinjector prompted  the exchange of the gun cavity.

\begin{figure}[!htb]
   \centering
   \includegraphics*[width=0.47\textwidth]{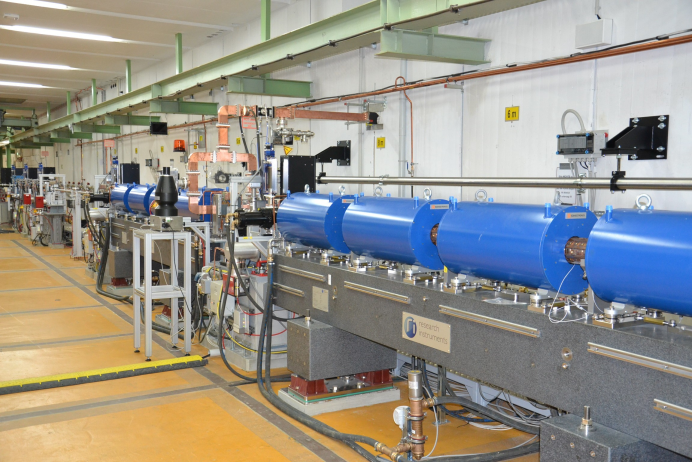}
   \caption{Finished setup of the ARES linac section.}
   \label{fig:ARESLinac}
\end{figure}

After the exchange of the gun with one with an upgraded design and the subsequent RF conditioning, the electron beam properties will be measured at the exit of the RF photoinjector again. As a next step, it is planned to guide the beam through the two travelling wave structures of the ARES linac section. Here, the beam can be accelerated up to 155\,MeV. The infrastructure of the linac is well prepared. The two linac cavities were fully conditioned during the last months. All magnets, including 8 solenoids around the travelling wave structures, several quadrupoles and steerer magnets, are installed. All diagnostics such as screen stations and a toroid for charge measurements are part of the control system and ready for operation. Figure~\ref{fig:ARESLinac} shows the complete setup of the ARES linac section.
\\ 
In the upcoming months, we aim to continue the commissioning of the ARES linac. The parameters of the high energy electron beam must be determined after the linac. First iterations to minimize the bunch length via velocity bunching are foreseen. Furthermore, the commissioning of the experimental area is expected for the last quarter of 2020 \cite{Burkart2019}. In this area, we will start the test of dielectric structures as ultra-compact particle accelerators in the context of the ACHIP collaboration \cite{England2014,Peralta2013,Breuer2013,Mayet2018}. The setup of the experimental chamber, including the ACHIP experiment, is finished and ready for operation.

Beginning in 2021, the installation of a bunch compressor \cite{Lemery2019} and the Polarix TDS \cite{Marx2019} will bring ARES closer to the goal of generating and characterizing fs to sub-fs short electron bunches. New diagnostics, such as the low-charge, beam profile monitor STRIDENAS will be tested in the future \cite{Jaster_Merz_2020}.


\section{Acknowledgements}

The authors acknowledge the work of all the DESY technical groups involved in planning, construction and operation of the ARES facility. Our special thanks go to Sven Lederer, Jakob Hauser (MVS group), Stefan Baark, Olaf Rasmussen (MEA group), Markus Hüning, Ingo Peperkorn, Jörg Herrmann (MIN group), Sven Pfeiffer, Mikheil Titberidze (MSK group), Lutz Winkelmann, Caterina Vidoli (FSLA group), Gero Kube, Artem Novokshonov, Matthias Werner, Gunnar Priebe (MDI group), Tim Wilksen, Olaf Hensler (MCS group), Christian Helwich, Bastian Belusic (MKK group), Klaus Flöttmann, Reinhard Brinkmann (MPY group).\\
   
In addition, we would like to mention our collaboration partners at ATHENA and ACHIP. The Helmholtz Association funds the ATHENA project. ACHIP related activities are partially funded by the Gordon and Betty Moore foundation (GBMF4744). Access via the ARIES-TNA program will partially be sponsored via the European Unions Horizon 2020 Research and Innovation programme under grant agreement No 730871. The AXSIS project has received funding from the European Research Council under the European Union’s Seventh Framework Programme (FP/2007-2013) / ERC Grant Agreement n. 609920’.

\section*{References}

\bibliographystyle{elsarticle-num}
\bibliography{\jobname}

\end{document}